# Anisotropic heat conduction of coherently transported phonons in InGaO$_3$(ZnO)$_m$ single crystal films with superlattice structures

Hai Jun Cho,[1]* Yuzhang Wu,[2] Youngha Kwon,[3] Jiajun Qi,[4] Yuna Kim,[1] Keiji Saito,[5] and Hiromichi Ohta[1]

[1]Research Institute for Electronic Science, Hokkaido University, N20W10, Kita, Sapporo 001-0020, Japan

[2]Graduate School of Information Science and Technology, Hokkaido University, N14W9, Kita, Sapporo 060-0814, Japan

[3]School of Engineering, Hokkaido University, N13W8, Kita, Sapporo 060-8628, Japan

[4]Graduate School of Life Science, Hokkaido University, N10W8, Kita, Sapporo 060-0811, Japan

[5]Department of Physics, Keio University, 3-14-1 Hiyoshi, Kohoku, Yokohama 223-8522, Japan

**Abstract**

Superlattices provide a great platform for studying coherent transportation of low-frequency phonons, which are the main issues in mastering the manipulation of heat conduction. Studies have shown that the dominating characteristics in the thermal conductivity of superlattice can be adjusted between wave-like and particle-like phonon properties depending on the superlattice period. However, the phonon coherence length and the phonon mean free path from Umklapp processes have not been defined in one superlattice system, and the transition from wave-like and particle-like behavior is not clear to date despite the extensive research efforts. In this study, we use $InGaO_3(ZnO)_m$ ($m$ = integer) single crystal films with superlattice structure to experimentally characterize the phonon coherence length as well as the Umklapp mean free path in one system. According to the results, the nature of heat conduction in superlattice can change in three different ways depending on the ratio between the phonon coherence length and the superlattice period. We also discuss the role of the phonon characteristic lengths in the heat conduction of superlattices and its anisotropy.

Heat is the most common form of energy, but it is mostly regarded as waste due to challenged associated with manipulating thermal energy. The main issue is the fact that thermal energy is associated with any forms of states capable of exhibiting kinetic energy, including phonons. Since phonons are bosons, controlling the overall phonon transportation requires accessing the entire phonon ensemble at a wide frequency range. This is much more difficult as to controlling electrical energy, for example, which only requires managing free charge carriers near the Fermi level.

In most cases, phononic heat conduction is dominated by high-frequency phonons that are diffusely scattered by crystal defects and anharmonicity in the lattice [1-6], which can be controlled reasonably well with dopants and grain boundaries. Contributions from low-frequency phonons is not significant due to their low energy, but these phonons coherently propagate through the lattice and have long propagation lengths [1, 7, 8]. As such, they are not sensitive to crystal defects and therefore, extremely difficult to engineer. This is a major concern for mastering the manipulation of heat conduction in solids, and understanding the coherent transportation of low-frequency phonons is of great interest in thermal management technology. However, creating a platform for studying coherent heat conduction is a challenge itself since diffuse scattering processes usually dominate the macroscopic heat conduction in solid.

Superlattices gathered enormous research attention in this regard, as they can exhibit very high densities of interfaces, which can significantly hinder the propagation of diffusely scattered phonons and allow coherent transport mechanism across interfaces to dominate the overall heat conduction [1, 7-10]. Thus, there have been extensive research efforts to clarify the heat conduction in superlattices. Theoretical studies provided several critical characteristics of coherent heat conduction in superlattices, such as group velocity changes associated with the phonon band folding and thermal conductivity ($\kappa$) as a function of the superlattice period ($d_{SL}$) [11-14]. However, there is a critical shortcoming in previous research efforts. Phonon coherence length ($\lambda$) and mean free path from Umklapp processes are the key characteristic lengths for the interpretation of heat conduction in superlattices. Although these two characteristic lengths are fundamentally different, they are not clearly distinguished within one material system in both theoretical and experimental studies [7, 9, 11, 12, 15]. Some experimental results from superlattices even showed signs of incoherent heat conduction [8, 16-18], and only a few experimental studies succeeded in partially demonstrating the theoretical concepts established for the coherent heat conduction in superlattices.

One possible explanation is the fact that most experimental studies were conducted on artificial superlattices, which can have additional anharmonicity associated with strains and defects from the lattice misfits of the superlattice constituents or the fabrication method [14, 16-18]. Natural superlattices were also investigated in polycrystalline forms, but the results were affected by randomly oriented grains although some specimens exhibited a strong preferred orientation [19, 20]. In this regard, the defect-free natural superlattice structure observed from $InGaO_3(ZnO)_m$ ($IGZO_m$) single crystalline films can potentially offer a solution. This material system consists of $GaO(ZnO)_m^+$ blocks separated by pristine $InO_2^-$ interfaces [21-23]. As the $d_{SL}$ and orientations of $IGZO_m$ films can be changed even in the single crystalline forms [10], this can be a perfect platform for investigating the fundamental characteristic lengths associated with the heat conduction in superlattices. In this study, we address the $\kappa$ of single crystalline $IGZO_m$ films at wide ranges of temperature and $d_{SL}$. The results show that the nature of heat conduction in superlattice is determined by the ratio between $\lambda$ and $d_{SL}$ (**Figure 1**). We experimentally demonstrate several essential concepts in superlattices, including $\lambda$, phonon mean free paths, anisotropic heat conduction, and vibrational coupling of superlattice constituents. We note here that the term 'coherent' in the context of this manuscript refers to phonon transport mechanism. This should not be confused with 'coherent phonons', which refer to the collective in-phase atomic motions in solid.

$IGZO_m$ single crystalline films with $m$-values from 1 to 29 were fabricated by the reactive solid-phase epitaxy (R-SPE) method on yttria-stabilized zirconia (YSZ) single crystal substrates, which is described elsewhere [10, 21-23]. Since the $[000h]$ axis of $IGZO_m$ grows parallel to YSZ [111] axis, (111)-oriented YSZ substrates were used for normally oriented

IGZO$_m$ while (110)-oriented YSZ substrates were used to fabricate IGZO$_m$ films tilted by ~30°[22]. The observed $\kappa$ from the inclined films ($\kappa_{incl.}$, **Figure S1**) can be decomposed into the out-of-plane $\kappa$ ($\kappa_\perp$) from normally oriented IGZO$_m$ and in-plane $\kappa$ ($\kappa_\parallel$) if the superlattice tilt angle ($\theta$) is known [24-26]:

$$\kappa_{incl.} = \sqrt{\kappa_\perp^2 cos^2\theta + \kappa_\parallel^2 sin^2\theta}$$

The $\kappa_\parallel$ values were extracted from $\kappa_{incl.}$ and $\kappa_\perp$.

High-resolution X-ray diffraction (HRXRD, Cu Kα$_1$, ATX-G, Rigaku Co.) was performed to characterize the thickness, $d_{SL}$, $m$-value, and the average tilt angle of the IGZO$_m$ films on YSZ (110) substrates. Thickness can be an important parameter for the $\kappa$ of superlattices, and it has been experimentally demonstrated that the $\kappa$ of superlattices can be proportional to the thickness up to 9·$d_{SL}$ if the heat conduction is dominated by coherent phonon transportation [1]. Therefore, to eliminate possible thickness dependence on the $\kappa$ of IGZO$_m$ films, the thicknesses of the films were adjusted to be much greater than 9·$d_{SL}$, especially at low values of $d_{SL}$ (**Table S1**). Atomic force microscopy (AFM) was used to observe the surface morphology of the films. The $\kappa$ of the IGZO$_m$ films perpendicular to the surface was measured using time-domain thermoreflectance (TDTR, PicoTR, PicoTherm Co.[27]). Mo films were sputtered on the IGZO$_m$ films as the transducer. The obtained thermoreflectance signals were analyzed with a software package provided by the manufacturer. The confirmation of defect-free interfaces and pristine atomic structure in IGZO$_m$ superlattice films can be found in our previous study [10]. The Raman spectra of the IGZO$_m$ in polycrystalline forms were measured using inVia Reflex Micro-Raman spectroscope (Renishaw).

The structural characterizations of the films are shown in **Figure 2**. Periodically spaced peaks from (000$h$) planes according to the $d_{SL}$ can be observed from both the normally oriented and inclined IGZO$_m$ films. Atomically smooth step and terrace surfaces are seen from normally oriented IGZO$_m$ films, whereas stripe-like patterns are seen from the surface of inclined IGZO$_m$ films. The $\kappa_\perp$ of (000$h$)-oriented IGZO$_m$ films are plotted against $d_{SL}$ in **Figure 3a**, which shows a typical V-shaped pattern frequently observed from superlattices [7, 9, 11]. This is attributed to the change in the nature of phonon transportation process dominating the overall heat conduction. Phonon wavepackets in solids maintain their phases within $\lambda$. If $d_{SL} > \lambda$, phonons diffusely scattered at interfaces can behave like particles and propagate between interfaces, which allow them to contribute to the overall heat conduction. Therefore, decreasing the $d_{SL}$ decreases the phonon mean free path ($l$) and $\kappa_\perp$. This regime is referred to as incoherent since the heat conduction is dominated by diffusely scattered high-frequency acoustic and optical phonons, which change their phases upon scattering. In contrast, if $d_{SL} < \lambda$, the phase of the phonon wavepackets must be preserved across at least two interfaces, and diffusely scattered high-frequency phonons cannot pass interfaces. In this case, the overall heat conduction is dominated by low-frequency acoustic phonons with long wavelengths that can preserve their phases across interfaces. This regime is referred to as coherent since the heat conduction is dominated by phonons that coherently go through interfaces. Therefore, this regime is strongly affected by the wave-like characteristics of phonons, and reducing $d_{SL}$ increases $\kappa_\perp$ since it increases the phonon group velocity [7, 11, 28]. On the other hand, $\kappa_\parallel$ does not show a strong dependence on $d_{SL}$, except between $d_{SL} = 2.4$ nm and 4.2 nm (**Figure 3b**). Detailed analysis is a bit difficult here since $\kappa_\parallel$ values were calculated from two measured values and have large errors, but $\kappa_\parallel$ is also likely affected by the ratio between $d_{SL}$ and the phonon characteristic lengths, which can be obtained from $\kappa_\perp$.

The transition from incoherent regime to coherent regime in the out-of-plane heat transportation of IGZO$_m$ films occurs when $d_{SL}$ becomes comparable to $\lambda$. Therefore, the minimum in the V-shaped pattern (**Figure 3a**) indicates the average phonon coherence length, which turns out to be ~1.6 nm from the $d_{SL}$ of IGZO$_4$. With decreasing temperature, the depth of the V-pattern increases while the location of the minimum ($\equiv \lambda$) remains unchanged. Previous theoretical studies are consistent with changes in the depth of V-pattern [11, 12] but different on the location of minimum. Some studies predict that lowering temperature shifts the minimum to lower $d_{SL}$ values due to longer $l$ in the superlattice constituents [11, 12], but another study indicates the location of minimum does not depend on temperature [14]. In the present study, the location of minimum in the $\kappa_\perp$ of IGZO$_m$ does not significantly change with temperature.

Since phonons are the eigenstates of a harmonic potential, $\lambda$ and $l$ are strongly influenced by anharmonicity. Therefore, $\lambda$ and the location of $\kappa_\perp$ minimum should be related to the temperature dependence of the anharmonicity in the lattice, which can be addressed with the coefficient of thermal expansion (CTE). According to a previous study, the CTE of IGZO$_m$ does not depend on temperature nor the $m$-values below 923 K [29]. This indicates that the anharmonicity in IGZO$_m$ does not strongly depend on temperature, which is consistent with our results. While the abovementioned discrepancy between theoretical results on the location of $\kappa_\perp$ minimum can be attributed to the difference in the anharmonic terms in the interatomic potentials, there is still a small ambiguity:

previous theoretical studies relate the $\kappa_\perp$ minimum with the Umklapp mean free path without any mentioning of $\lambda$, although they clearly indicate that the V-shaped pattern is attributed to a crossover between particle-like and wave-like phonon transportation [11, 12, 14]. To provide clarity, it is necessary to extract the $l$ associated with the Umklapp processes in GaO(ZnO)$_m$$^+$ blocks ($l_{GZO}$) in our IGZO$_m$ system to distinguish $\lambda$ from $l_{GZO}$.

$l_{GZO}$ can be estimated from the temperature dependent $\kappa_\perp$. In the incoherent regime ($d_{SL}$ > $\lambda$), the out-of-plane $l$ of diffusely scattered phonons in the IGZO$_m$ films ($l_\perp$) can be expressed with $l_{GZO}$ and $d_{SL}$ using Matthiessen's rule:

$$\frac{1}{l_\perp} = \frac{1}{l_{GZO}} + \frac{1}{d_{SL}} \quad \text{if} \quad l_{GZO} < d_{SL}$$

$$\frac{1}{l_\perp} = \frac{1}{d_{SL}} \quad \text{if} \quad l_{GZO} > d_{SL}$$

While $d_{SL}$ is not affected by temperature, $l_{GZO}$ is affected by Umklapp processes and changes with temperature. Since $\kappa_\perp = \frac{1}{3}Cvl_\perp$ ($C$ = heat capacity, $v$ = the average phonon speed), the temperature dependence in $\kappa_\perp$ disappears if $l_{GZO} > d_{SL}$. **Figure 4a** shows the $\kappa_\perp$ of our IGZO$_m$ films from 173 K to 573 K. At high values of $d_{SL}$ ($m$ = 29), $\kappa_\perp$ monotonically decreases with increasing temperature due to reductions in $l_{GZO}$. This temperature dependence slowly diminishes with decreasing $d_{SL}$ and disappears at $d_{SL}$ = 3.2 nm ($m$ = 10), which suggests that $l_{GZO}$ is around 3.2 nm in this temperature range. This clearly distinguishes $l_{GZO}$ from $\lambda$ in IGZO$_m$, which demonstrates that the transition from incoherent to coherent heat conduction across superlattices is separated by the coherency of phonons while the temperature dependence is related to the mean free path.

Interestingly, the temperature dependence comes back at very low $d_{SL}$ values ($m$ = 1, 2)

(**Figure 4b**). We believe this is attributed to new vibrational states from the coupling of GaO(ZnO)$_m{}^+$ and InO$_2{}^-$. If the GaO(ZnO)$_m{}^+$ blocks become very thin, the system cannot only be regarded as GaO(ZnO)$_m{}^+$ blocks separated by InO$_2{}^-$ layers. Once the volume ratio of the superlattice constituents (GaO(ZnO)$_m{}^+$ and InO$_2{}^-$) approaches 1:1 at low $d_{SL}$ values, the density of states of the coupled vibrational mode will extend to the entire superlattice. The new coupled vibrational state will be able to travel through the superlattice and exhibit temperature dependance associated with the Umklapp processes. The Raman spectra of IGZO$_m$ confirm this hypothesis (**Figure S2**), as an additional mode from the coupling emerge at ~700 cm$^{-1}$ at low $m$-values. This is consistent with the Raman spectra of In$_2$O$_3$(ZnO)$_m$ superlattices [30], which have very similar crystal structure with IGZO$_m$.

Temperature dependent $\kappa_\parallel$ values are shown in **Figure 4c**. At the same $m$-values, $\kappa_\parallel$ was greater than $\kappa_\perp$ and exhibited stronger temperature dependence. Similar to $\kappa_\perp$, the temperature dependence disappears as $d_{SL}$ decreases. At low values of $d_{SL}$, small recessions in $\kappa_\parallel$ can be observed with increasing temperature, especially after 473 K. This may be attributed to Umklapp scattering of coupled vibrational modes, but we would like to note that the present $\kappa_\parallel$ results only confirm a strong hinderance in the temperature dependence since the range of recession is within the uncertainties (~20%). The sudden drop in $\kappa_\parallel$ observed from $m = 14$ to $m = 10$ can be explained using the Raman spectra again. As the $m$-value decreases, the intensity of ZnO-like mode at ~450 cm$^{-1}$ decreases, and a shoulder peak develops on its left, which indicates phonon localization [30]. This phenomenon becomes strongly evident at $m = 10$, where the $d_{SL}$ approximately matches $l_{GZO}$. Therefore, the drop in $\kappa_\parallel$ at $m = 10$ can be attributed to additional phonon scatterings from the localized modes, which explains the strong $d_{SL}$ dependence near $m = 10$ shown in **Figure 3b**.

If $d_{SL}$ becomes comparable to $\lambda$, $\kappa_\parallel$ is no longer significantly affected by $d_{SL}$. This indicates a change in the dominating heat conduction mechanisms along the in-plane direction from diffusive scattering to specular scattering, which is predicted from theoretical studies [15]. The initial reduction in $\kappa_\parallel$ with decreasing $d_{SL}$ is attributed to the phonon scatterings at interfaces. However, if $\lambda > d_{SL}$, phonon wavepackets will always be in physical contact with interfaces, and the interfacial scattering rate will not strongly depend on temperature nor $d_{SL}$. At $d_{SL} = 1.1$ nm ($m = 2$), $\kappa_\parallel$ slightly increases from $d_{SL} = 1.6$ nm ($m = 4$), which is likely attributed to the coupled vibrational states seen earlier from the temperature dependent $\kappa_\perp$ (**Figure 4a**) and the Raman spectra of IGZO$_m$ (**Figure S2**).

This study identified two important phonon characteristic lengths in IGZO$_m$ superlattices: $\lambda$ and $l_{GZO}$, which control the dominating heat conduction mechanism. If $d_{SL} \geq \lambda$, both the $\kappa_\perp$ and $\kappa_\parallel$ of IGZO$_m$ are mainly attributed to the particle-like behavior of high-frequency phonons. Depending on the ratio between $d_{SL}$ and $l_{GZO}$, the $\kappa_\perp$ of IGZO$_m$ can be dominated by temperature-dependent Umklapp scattering processes ($d_{SL} > l_{GZO}$) or temperature-independent diffuse scattering at InO$_2^-$ interfaces ($d_{SL} < l_{GZO}$). The $d_{SL}$ and $l_{GZO}$ also play an important role in the $\kappa_\parallel$ of IGZO$_m$, which sharply drops with decreasing $d_{SL}$ if $d_{SL} \sim l_{GZO}$ due to phonon localizations. Once $d_{SL}$ becomes shorter than $\lambda$, the diffusive phonon characteristics disappear from the macroscopic heat transportation. In this regime, coherent phonon transportation dominates $\kappa_\perp$, and decreasing $d_{SL}$ enhances $\kappa_\perp$. However, at very small values of $d_{SL}$, the vibrational coupling of GaO(ZnO)$_m^+$ and $InO_2^-$ restores some diffusive nature in $\kappa_\perp$. In addition, the phonon wavepackets in this regime ($d_{SL} < \lambda$) are always in contact with $InO_2^-$ interfaces, and the main scattering mechanism for $\kappa_\parallel$ becomes specular, which does not strongly depend on temperature nor $d_{SL}$. Simple analogy of the phonon characteristic lengths addressed in the present study would be the thermal de Broglie wavelength ($\sim\lambda$) and mean free path of electrons ($\sim l_{GZO}$), which represent wave-like and particle-like behavior of electrons, respectively. We believe our results offer a fundamental contribution to knowledge in better understanding the heat conduction in superlattices.

**Data availability**

The data that support the findings of this study are available from the corresponding authors upon reasonable request.

**Acknowledgements**

This research was supported by Grants-in-Aid for Innovative Areas (19H05791) from the Japan Society for the Promotion of Science (JSPS). H.J.C. acknowledges the support from Nippon Sheet Glass Foundation for Materials Science and Engineering. The support from China Scholarships Council (Y.W.: 201908050162) is also greatly appreciated. H.J. Cho and Y. Wu contributed equally to this work.

**Competing financial interests**

The authors declare no competing interests.

**Author Contributions**

Sample fabrications: Y. Wu, H.J. Cho, Y. Kwon;

Structural characterizations and TDTR: Y. Wu, H.J. Cho, Y. Kwon

Raman spectroscopy: J. Qi, Y. Kim

Data analysis: H.J. Cho, K. Saito

Manuscript preparation: H.J. Cho, Y. Wu

Principal investigation: H. Ohta

**Corresponding author**

Correspondence and requests for materials should be addressed to:

H.J.C. (joon@es.hokudai.ac.jp)

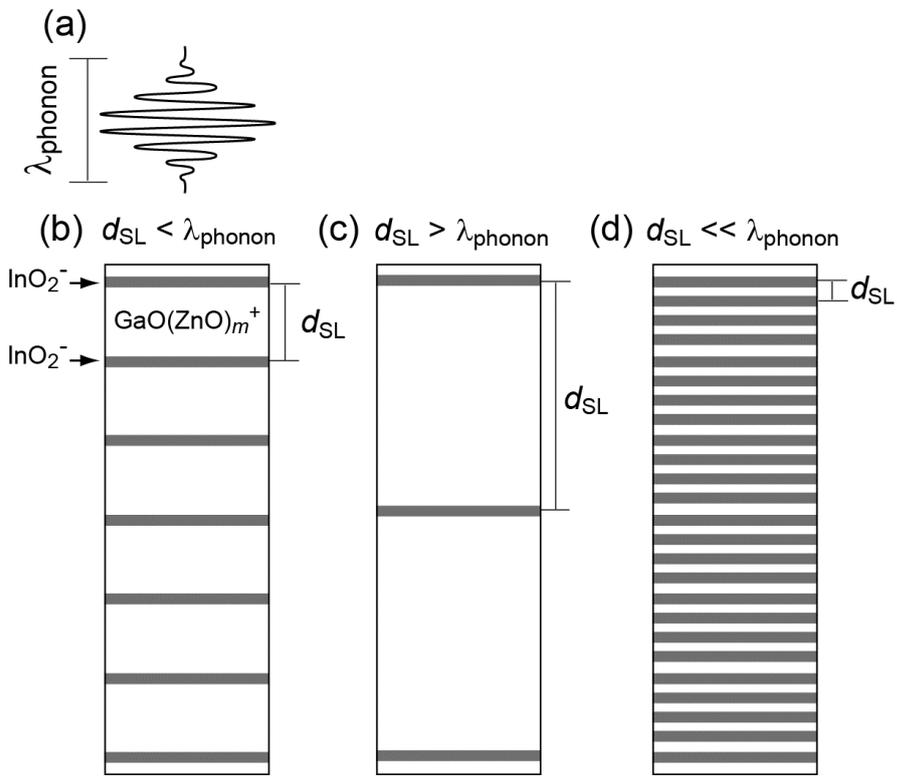

**Figure 1** | Ratio of characteristic lengths in single-crystalline IGZO$_m$ films that determine the nature of heat conduction.

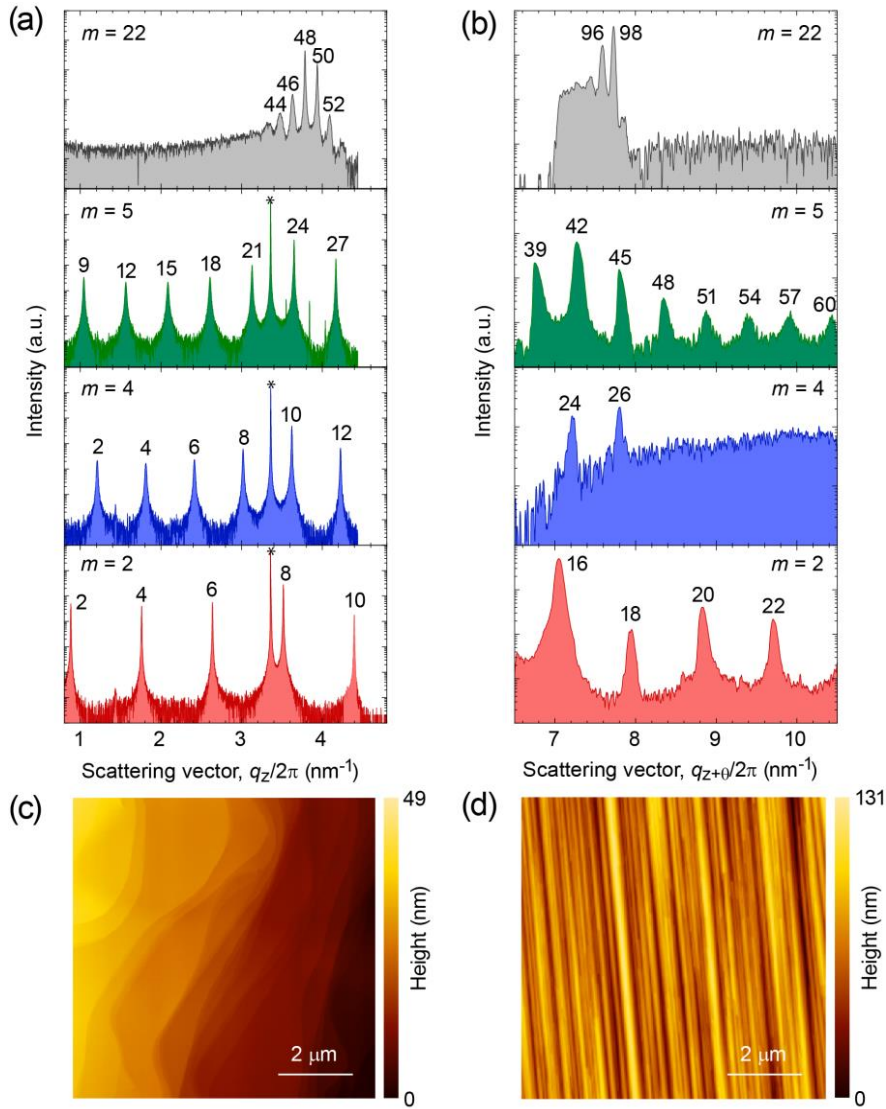

**Figure 2** | XRD patterns of the IGZO$_m$ films grown on (a) (111) YSZ (*) and (b) (110) YSZ. Periodic peaks from (000$h$) planes according to the changes in the $m$-values and the respective $d_{SL}$ can be seen. The measured tilt angles on the superlattice on (110) YSZ turned out to be 31.5°, which is close to the difference between YSZ (110) and (111) (30°). (c, d) Topographic AFM images of the IGZO$_m$ films grown on (c) (111) YSZ and (d) (110) YSZ. Atomically smooth stepped and terraced surface is seen in (c), while stripe patterns are seen in (d).

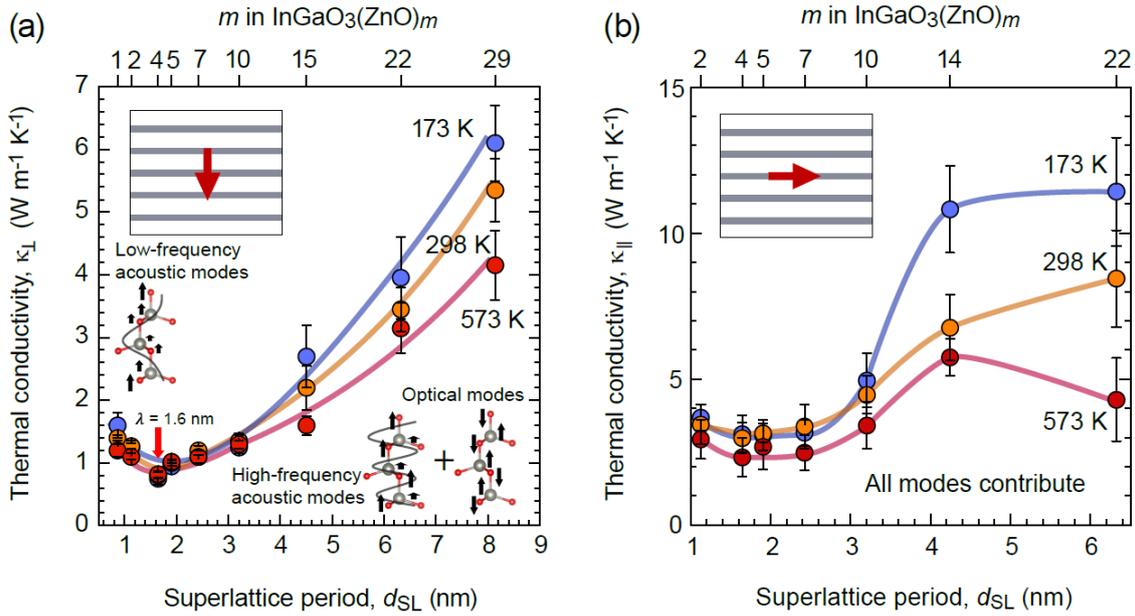

**Figure 3** | (a) $\kappa_\perp$ of IGZO$_m$ films as a function of $d_{SL}$. The crossover from coherent to incoherent heat conduction can be observed at $m = 4$, suggesting a coherence length of ~1.6 nm. Examples of phonon motions dominating coherent and incoherent regime are shown. More phonon motions can be found in the supplementary [31]. (b) $\kappa_\parallel$ of IGZO$_m$ films as a function of $d_{SL}$. The overall pattern has a stronger temperature dependence compared to that along the out-of-plane direction, which is reasonable since the interfaces are parallel to heat propagation. Interestingly, strong dependence on $d_{SL}$ is only observed between 2.4 nm and 4.2 nm.

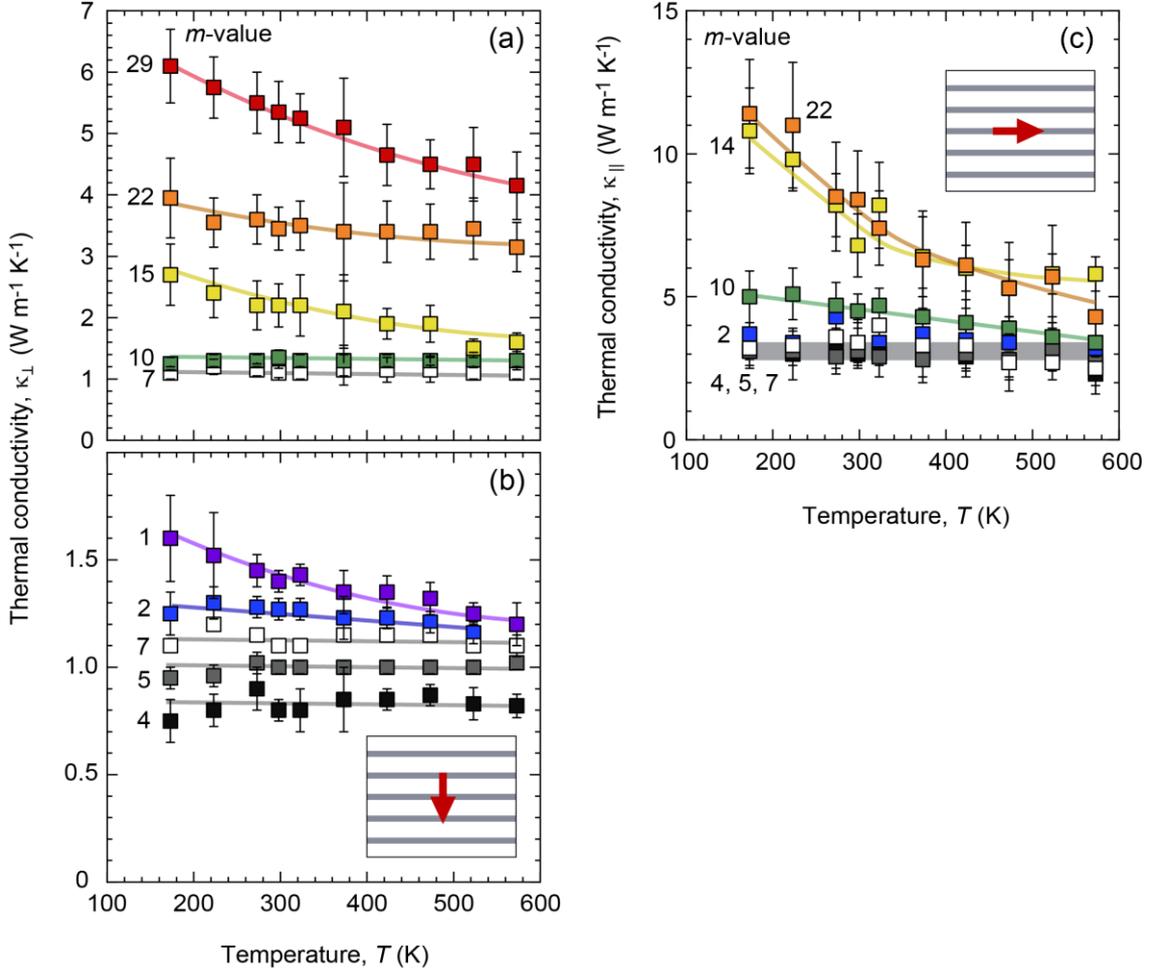

**Figure 4** | Temperature dependent thermal conductivity of the IGZO$_m$ films. (a) $\kappa_\perp$ from $m = 29$ to $m = 7$. At high $m$-values, $\kappa_\perp$ decreases with increasing temperature, which is similar with ZnO single crystals (**Figure S3**). This is a typical behavior from Umklapp processes, which dominate coarse-grained bulk materials. The temperature dependence slowly disappears at $m = 10$, indicating that $d_{SL}$ approached $l_{GZO}$. This clearly distinguishes $\lambda$ from $l_{GZO}$ and demonstrates the coherent-incoherent crossover is only associated with $\lambda$. (b) $\kappa_\perp$ from $m = 7$ to $m = 1$. Initially, there is no temperature dependence, but it appears at very low values of $m$. This is likely associated with the coupling of GaO(ZnO)$_m^+$ blocks and InO$_2^-$ interfaces. (c) $\kappa_\parallel$ from $m = 22$ to $m = 2$. If $d_{SL} > \lambda$, the temperature dependence is strong like ZnO. However, if $d_{SL} < \lambda$, the temperature dependence drastically decreases.

# Anisotropic heat conduction of coherently transported phonons in InGaO$_3$(ZnO)$_m$ single crystal films with superlattice structures

Hai Jun Cho,[1]* Yuzhang Wu,[2] Youngha Kwon,[3] Jiajun Qi[4], Yuna Kim[1], Keiji Saito[5], and Hiromichi Ohta[1]

[1]Research Institute for Electronic Science, Hokkaido University, N20W10, Kita, Sapporo 001-0020, Japan

[2]Graduate School of Information Science and Technology, Hokkaido University, N14W9, Kita, Sapporo 060-0814, Japan

[3]School of Engineering, Hokkaido University, N13W8, Kita, Sapporo 060-8628, Japan

[4]Graduate School of Life Science, Hokkaido University, N10W8, Kita, Sapporo 060-0811, Japan

[5]Department of Physics, Keio University, 3-14-1 Hiyoshi, Kohoku, Yokohama 223-8522, Japan

**Table S1** – Thicknesses of IGZO$_m$ films

| Normally oriented IGZO$_m$ films (on YSZ 111) ||
|---|---|
| $m$-value | Thicknesses (nm) |
| 1 | 268 |
| 2 | 278 |
| 4 | 143 |
| 5 | 177 |
| 7 | 217 |
| 10 | 223 |
| 15 | 152 |
| 22 | 113 |
| 29 | 90 |
| **Inclined oriented IGZO$_m$ films (on YSZ 110)** ||
| $m$-value | Thicknesses (nm) |
| 2 | 169 |
| 4 | 112 |
| 5 | 102 |
| 7 | 130 |
| 10 | 130 |
| 14 | 66 |
| 22 | 98 |

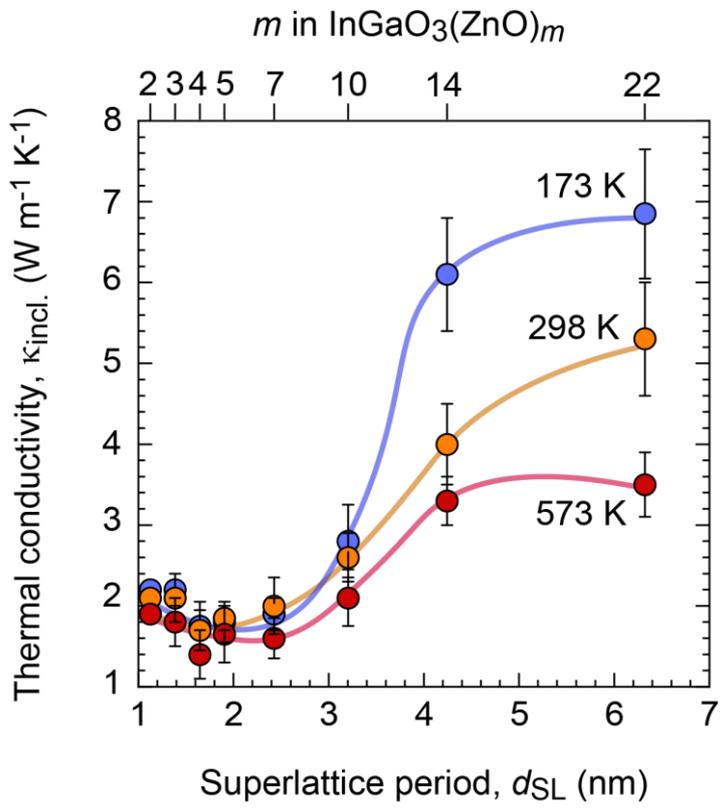

**Figure S1** | Observed thermal conductivities $\kappa_{incl.}$ from the inclined IGZO$_m$ films on (110) YSZ substrates.

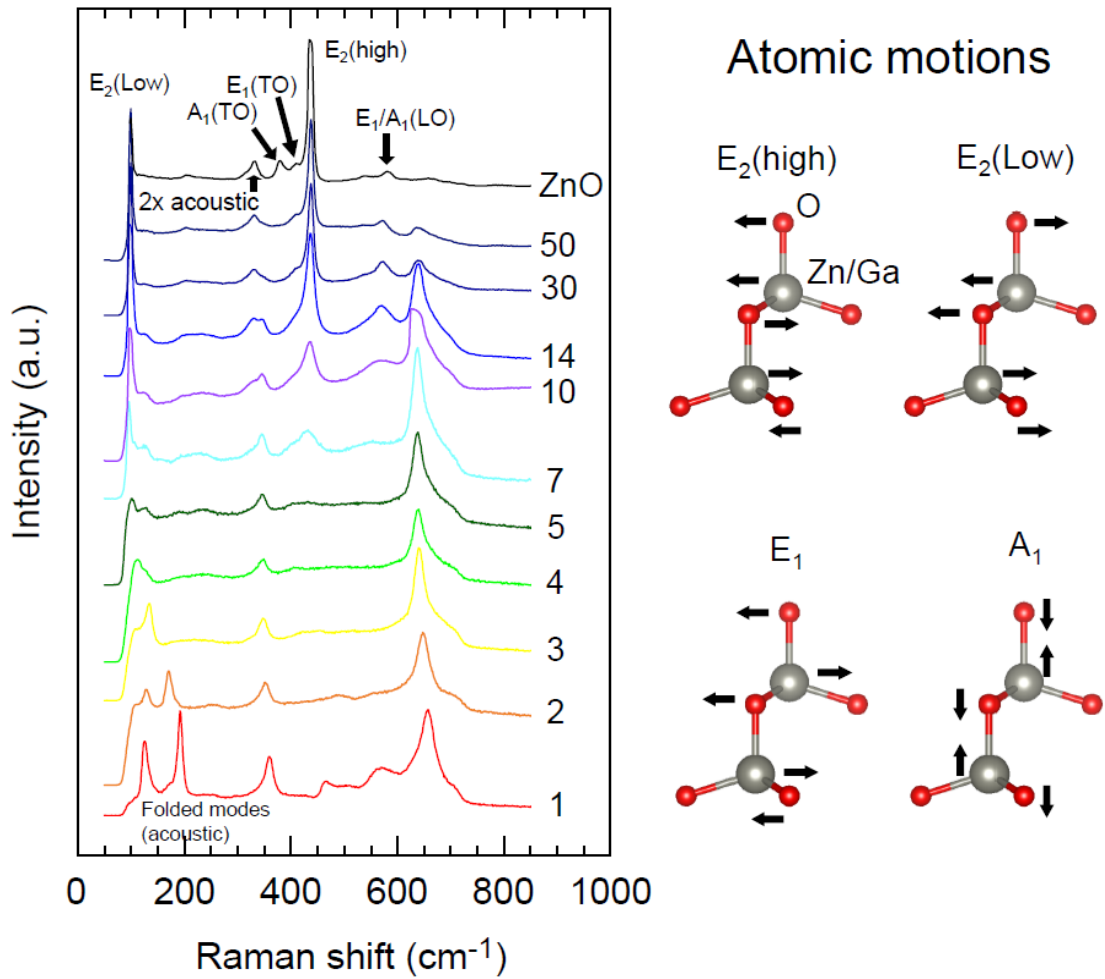

**Figure S2** | Raman spectra of IGZO$_m$ with different $m$-values and atomic motions of the observed vibrational modes [30,31]. TO and LO refer to transverse optical and longitudinal optical modes, respectively. With decreasing $m$-values (decreasing $d_{SL}$), new modes emerge at ~700 cm$^{-1}$ due to the coupling of GaO(ZnO)$_m{}^+$ and InO$_2{}^-$. The other new modes at < 200 cm$^{-1}$ are attributed to band folding. On the other hand, the ZnO-like mode at ~450 cm$^{-1}$ disappears and develops a shoulder peak at its left with decreasing $m$-values. According to ref. 30, this implies phonon localization.

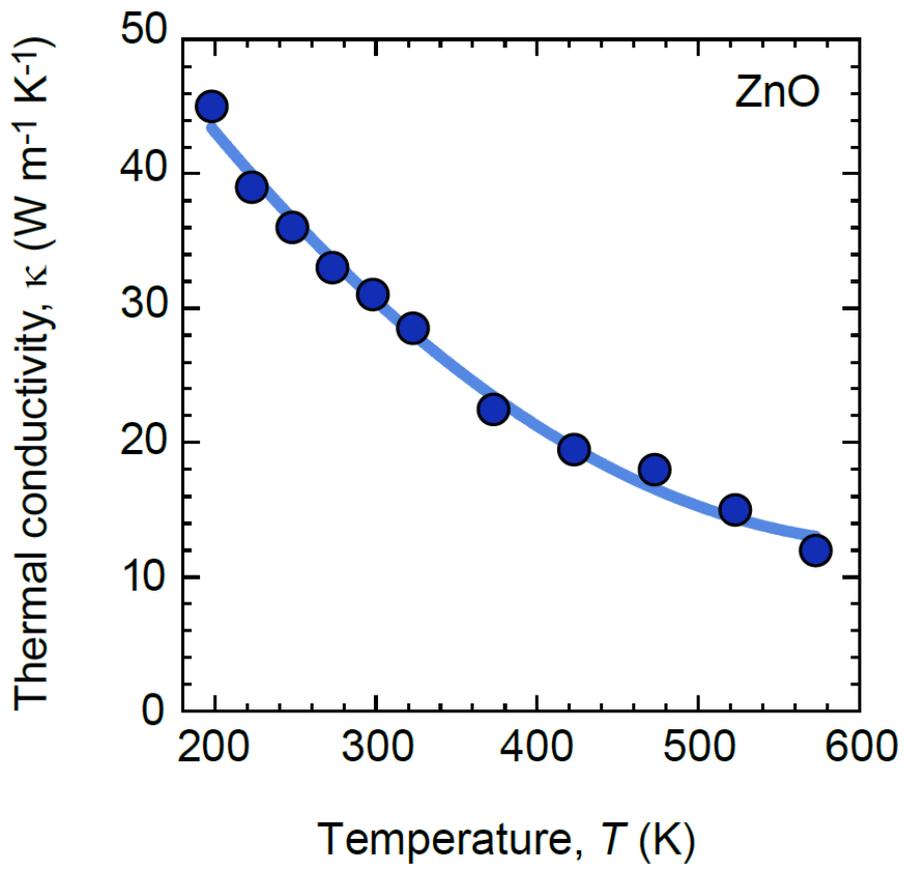

**Figure S3** | Temperature dependence of thermal conductivity for *c*-axis oriented ZnO single crystal.